\documentstyle[12pt]{article}
\def\beqa{\begin{eqnarray}}
\def\eeqa{\end{eqnarray}}
\def\beq{\begin{equation}}
\def\eeq{\end{equation}}
\def\disp{\displaystyle}

\def\vol{d^4x\,\sqrt{-g}}

\def\half{\frac{1}{2}}
\def\gu{g^{\mu\nu}}
\def\gd{g_{\mu\nu}}
\def\gbru{{\bar{g}}^{\mu\nu}}
\def\gbrd{{\bar{g}}_{\mu\nu}}

\def\pa{\partial}

\def\al{{\alpha}}
\def\be{{\beta}}
\def\gam{{\gamma}}
\def\del{{\delta}}

\def\lam{{\lambda}}

\def\ome{{\omega}}
\def\GAM{{\Gamma}}
\def\LAM{{\Lambda}}
\def\OME{{\Omega}}

\def\da{_{\alpha}}

\def\umu{^{\mu}}

\def\dmunu{_{\mu\nu}}

\def\dmuunu{_{\mu}^{~\nu}}

\def\uab{^{\alpha\beta}}

\def\udea{_{;}^{\alpha}}

\def\ddemu{_{;\mu}}  
\def\ddenu{_{;\nu}}  
\def\ddea{_{;\alpha}}  
  
\def\ddemunu{_{;\mu\nu}}

\def\ddeab{_{;\alpha\beta}}

\def\L{{{\cal L}}}
\def\f{{\phi}}
\def\ka{{\kappa}}
\def\gbr{{\bar{g}}}
\def\ad{{\dot{a}}}
\def\add{{\ddot{a}}}
\def\ap{{a^{'}}}

\def\apq{{a^{'2}}}
\def\abr{{\bar{a}}}
\def\abrd{{\dot{\bar{a}}}}
\def\abrdd{{\ddot{\bar{a}}}}
\def\abrp{{{\bar{a}}^{'}}}

\def\abrpq{{{\bar{a}}^{'2}}}
\def\fd{{\dot{\phi}}}
\def\fdd{{\ddot{\phi}}}
\def\fp{{\phi^{'}}}

\def\fpq{{\phi^{'2}}}
\def\fbr{{\bar{\phi}}}
\def\fbrd{{\dot{\bar{\phi}}}}
\def\fbrdd{{\ddot{\bar{\phi}}}}
\def\fbrp{{{\bar{\phi}}^{'}}}

\def\fbrpq{{{\bar{\phi}}^{'2}}}
\def\VDEF{{V_{\phi}}}
\def\VBR{{\bar{V}}}
\def\VBRDEF{{\bar{V}_{\bar{\phi}}}}
\def\FDEF{{F_{\phi}}}
\def\FD{{\dot{F}}}

\def\tbr{{\bar{t}}}
\def\LBR{{\overline{L}}}
\def\EBR{{\overline{E}}}

\def\ie{{\it i.e. }}

\def\pr{{\it Phys. Rev.}\ }
\def\prl{{\it Phys. Rev. Lett.}\ }
\def\pl{{\it Phys. Lett.}\ }
\def\np{{\it Nucl. Phys.}\ }

\def\ijmp{{\it Int. Journ. Mod. Phys.}\ }
\def\ijtp{{\it Int. Journ. Theor. Phys.}\ }

\def\cqg{{\it Class. Quantum Grav.}\ }
\def\aph{{\it Ann. Phys. (N.Y.)}\ }

\def\prs{{\it Proc. R. Soc.}\ }
\def\grg{{\it Gen. Relativ. Grav.}\ }

\def\ncim{{\it Nuovo Cim.}\ }

\def\prep{{\it Phys. Rep.}\ }

\def\hpa{{\it Helv. Phys. Acta}\ }

\def\bib#1{$^{\ref{#1}}$}

\begin{document}
\def\bib#1{[{\ref{#1}}]}
\begin{titlepage}
\title{Some aspects of the cosmological conformal equivalence between
``Jordan Frame'' and ``Einstein Frame''} 

 \author{{S. Capozziello, R. de Ritis, A. A. Marino}
\\ {\em Dipartimento di Scienze Fisiche, Universit\`{a} di Napoli,}
\\ {\em Istituto Nazionale di Fisica Nucleare, Sezione di Napoli,}\\
   {\em Mostra d'Oltremare pad. 19 I-80125 Napoli, Italy.}}
	      \date{}
	      \maketitle
	      \begin{abstract}
The conformal equivalence between Jordan frame and Einstein frame can 
be used in order to search for exact solutions in general theories of 
gravity in which scalar fields are minimally or nonminimally coupled 
with geometry. In the cosmological arena a relevant role is played by
the time parameter in which dynamics is described. In this paper we
discuss such issues considering also if cosmological Noether
symmetries in the ``point--like'' Lagrangian are conformally
preserved. 

Through this analysis and through also a careful analysis of the
cosmological parameters $\OME$ and $\LAM$, it is possible to
contribute to the discussion on which is the physical system. 

	      \end{abstract}

\vspace{20. mm}
e-mail address: \\
 capozziello@axpna1.na.infn.it \\
 deritis@axpna1.na.infn.it \\
 marino@axpna1.na.infn.it
	      \vfill
	      \end{titlepage}

\section{\normalsize \bf Introduction}

Alternative theories of gravity has been formulated and investigated
in different context. The Brans--Dicke approach \bib{brans-dicke}, 
closely related to the Jordan approach \bib{jordan-fierz}, has been
carried on in the context of a generalization of the Mach's principle;
in that approach the Einstein theory of gravitation is modified by
introducing a scalar field with a non standard coupling with gravity,
\ie the gravitational coupling turns out to be no longer constant.
Later on more general couplings have been considered, and the
compatibility of such approaches with the different formulations of
the Equivalence Principle have been considered \bib{dicke} \bib{nmc}
\bib{damour1} \bib{damour2} \bib{qnmc}. 

A generalization of the standard gravity comes also from quantum field
theories on curved space--times; in such context we find the so called
higher order gravitational theories \bib{ordsup} \bib{qordsup}
\bib{birrel}. 

In all these approaches, the problem of reducing both these two kinds
of more general theories in Einstein standard form has been
estensively treated; one can see that, through a
``Legendre'' transformation on the metric, higher order theories,
under suitable regularity conditions on the Lagrangian, take the form
of the Einstein one in which a scalar field (or more than one) is the
source of the gravitational field (see for example \bib{ordsup}
\bib{sokolowski} \bib{magnano-soko}); on the other side, it has been
studied the equivalence between models with $G$--variable with the
Einstein standard gravity through a suitable conformal transformation
(see \bib{dicke} \bib{nmc}). 

In this paper we analyse, through an appropriately defined conformal
transformation, the problem of the equivalence between the non
minimally coupled (NMC) theories and the Einstein gravity for
scalar--tensor theories in absence of ordinary matter. First, we will
do it in the general context and then in the cosmological case, that
is, we will study the conformal invariance with the hypotheses of
homogeneity and isotropy. In such case we also consider the case in
which ordinary matter is present beside the scalar field and we do
some consideration on the problem of which is the ``physical system''
between the two conformally equivalent systems \bib{ordsup}
\bib{sokolowski} \bib{francaviglia}. 

Furthermore, we analyse the relation between the conformal equivalence
and the existence of a Noether symmetry in the $(a,~\f)$--space seen as
configuration space (\ie in the minisuperspace), where the 
cosmological ``point--like'' Lagrangian is defined (we will better
clarify the meaning of such expression in our forthcoming
considerations); of course such Lagrangian density comes from the
general field Lagrangian density once homogeneity and isotropy are
assumed \bib{nuovocim}. 

We conclude discussing some examples of physical interest. 

\section{\normalsize \bf Conformally equivalent theories}

In four dimension, the most general action involving gravity 
nonminimally coupled with one scalar field is
\beq
\label{1}
{\cal A}= \int \vol \left[F(\f) R+ \half \gu \f\ddemu \f\ddenu- V(\f)
\right] 
\eeq
where $R$ is the Ricci scalar, $V(\f)$ and $F(\f)$ are generic
functions describing respectively the potential for the field $\f$ and
the coupling of $\f$ with the gravity; the metric signature is
$(+---)$. We use Planck units.

The variation with respect to $\gd$ gives rise to the field equations
\beq
\label{2}
F(\f) G\dmunu= -\half T\dmunu- g\dmunu \Box_{\GAM} F(\f)+ F(\f)\ddemunu
\eeq
which are the generalized Einstein equations; here $\Box_{\GAM}$ is the
d'Alembert operator with respect to the connection $\GAM$; 
\beq
\label{3}
G\dmunu= R\dmunu- \half R \gd
\eeq
is the Einstein tensor, and
\beq
\label{4}
T\dmunu= \f\ddemu \f\ddenu- \half g\dmunu \f\ddea \f\udea+ g\dmunu 
V(\f)
\eeq
is the energy--momentum tensor relative to the scalar field. 

The variation with respect to $\f$ provides the Klein--Gordon equation
\beq
\label{5}
\Box_{\GAM} \f- R \FDEF(\f)+ \VDEF(\f)= 0
\eeq
where $\FDEF= dF(\f)/d\f$, $\VDEF= dV(\f)/d\f$. This last equation is
equivalent to the Bianchi contracted identity (see \bib{nuovocim}). 

Let us consider now a conformal transformation on the metric $\gd$ 
\bib{wald}, that is 
\beq
\label{10}
\gbrd= e^{2 \ome} \gd
\eeq
in which $e^{2 \ome}$ is the conformal factor. Under this 
transformation the connection, the Riemann and Ricci tensors, and 
the Ricci scalar transform in the corresponding way 
\bib{wald}, so that the Lagrangian density in (\ref{1}) becomes
\beq
\label{11}
\begin{array}{ll}
\sqrt{-g}(F R+ \half \gu \f\ddemu \f\ddenu- V) & = \sqrt{-\gbr}
e^{-2 \ome}(F \bar{R}- 6 F \Box_{\bar{\Gamma}} \ome+ \\ 
~ & ~ \\
~ & -6 F \ome\ddea \ome\udea+ \half \gbru \f\ddemu \f\ddenu- e^{-2
\ome} V) 
\end{array}
\eeq
in which $\bar{R}$, $\bar{\Gamma}$ and $\Box_{\bar{\GAM}}$ are
respectively the Ricci scalar and the connection relative to the
metric $\gbrd$, and the d'Alembert operator relative to the connection
$\bar{\GAM}$. If we require the theory in the metric $\gbrd$ to appear
as a standard Einstein theory, we get at once that the conformal
factor has to be related to $F$, that is (see also \bib{nmc} for a
review of particular cases) 
\beq
\label{12}
e^{2 \ome}= -2 F.
\eeq
We see that $F$ must be negative. Using this relation, the Lagrangian
density (\ref{11}) becomes 
\beq
\label{13}
\sqrt{-g} \left( F R+ \half \gu \f\ddemu \f\ddenu- V\right) =
\sqrt{-\gbr} \left( -\half \bar{R}+ 3  \Box_{\bar{\Gamma}} \ome+
\frac{3 \FDEF^2- F}{4 F^2} \f\ddea \f\udea- \frac{V}{4 F^2}\right), 
\eeq
Introducing a new scalar field $\fbr$ and the potential $\VBR$,
respectively, defined by 
\beq
\label{14}
\fbr\ddea= \sqrt{\frac{3\FDEF^2- F}{2 F^2}}\, \f\ddea,~~~ \VBR(\fbr(\f))= 
\frac{V(\f)}{4 F(\f)}
\eeq
we get
\beq
\label{15}
\sqrt{-g} \left( F R+ \half \gu \f\ddemu \f\ddenu- V\right) =
\sqrt{-\gbr} \left( -\half \bar{R}+ \half \fbr\ddea \fbr\udea-
\VBR\right), 
\eeq
which is the usual Einstein--Hilbert Lagrangian density plus the
standard Lagrangian density relative to the scalar field $\fbr$ (see
\bib{nuovocim}). (We have not considered the divergence--type term
appearing in the Lagrangian (\ref{15}); we will return on this point
in our forthcoming considerations). Therefore, any nonminimally
coupled theory, in absence of ordinary matter, is conformally
equivalent to an Einstein theory, being the conformal transformation
and the potential opportunely defined by (\ref{12}) and (\ref{14})
(see also \bib{damour1}). The converse is also true: for a given 
$F(\f)$, such that $3 \FDEF^2- F> 0$, we can trasform a standard 
Einstein theory into a NMC theory. This means that, in principle, if
we are able to solve the field equations in the framework of the
Einstein theory in presence of a scalar field with a given potential,
we should be able to get the solutions for the class of nonminimally
coupled theories, assigned by the coupling $F(\f)$, via the conformal
transformation defined by the (\ref{12}) the only constraint being the
second of (\ref{14}). This is exactly what we are going to discuss in
the cosmological context in cases in which the potentials as well as
the couplings are relevant from the point of view of the fundamental
physics. 

Following the standard terminology, we denote here as ``Einstein 
frame'' the framework of the Einstein theory, also indicated as 
minimally coupled theory and as ``Jordan frame'' the framework 
of the nonminimally coupled theory \bib{jordan-fierz}.

There are some remarks to do with respect to (\ref{13}) and
(\ref{14}): first we want to stress that the ``new'' scalar field as
defined in (\ref{14}) is given in differential form in terms of the
``old'' one and its integration can be not trivial; the second remark
concerns the divergence appearing in (\ref{13}). The transformed
Lagrangian density obtained from (\ref{11}) imposing (\ref{12})
contains a divergence term, in which appears not only the metric but
also its derivative, through the connection $\bar{\Gamma}$. Therefore
the equivalence of this total Lagrangian density to the
Einstein--Hilbert Lagrangian density plus scalar field is not trivial.
To check that they are actually equivalent, let us perform the
conformal transformation (\ref{10}) on the field Eqs. (\ref{2}),
obtaining 
\beq
\label{16b}
\begin{array}{ll}
{\bar{G}}\dmunu & = \disp{ \left( -\frac{1}{2 F}+ \frac{F_{\f \f}}{F}+
\frac{2 \ome_{\f} \FDEF}{F}- 2 {\ome_{\f}}^2- 2 \ome_{\f \f} \right)
\f\ddemu \f\ddenu+ } \\ 
~ & ~ \\
~ & \disp{ +\left( \frac{1}{4 F}- \frac{F_{\f \f}}{F}+ \frac{\ome_{\f}
\FDEF}{F}- {\ome_{\f}}^2+ 2 \ome_{\f \f} \right) \gbrd \f\ddea \f\udea+
\left( -\frac{F_{\f}}{F}+ 2 \ome_{\f}\right) \gbrd \Box_{\bar{\Gamma}}
\f+} \\ 
~ & ~ \\
~ & \disp{ +\left( \frac{F_{\f}}{F}- 2 \ome_{\f}\right)
(\nabla_{\bar{\Gamma}})_\mu (\nabla_{\bar{\Gamma}})_\nu \f- \frac{1}{2 F}
e^{-2 \ome} \gbrd V}
\end{array}
\eeq
in which $(\nabla_{\bar{\Gamma}})_\mu$ is the covariant derivative with 
respect to $x\umu$ relative to the connection $\bar{\Gamma}$. We see, 
from (\ref{16b}), that if $\ome$ satisfies the relation
\beq
\label{17b}
\frac{\FDEF}{F}- 2 \ome_\f= 0
\eeq
Eqs. (\ref{16b}) can be rewritten as
\beq
\label{18b}
{\bar{G}}\dmunu= \frac{3 \FDEF^2- F}{2 F^2} \f\ddemu \f\ddenu- \gbrd 
\frac{3 \FDEF^2- F}{2 F^2}  \f\ddea \f\udea- \gbrd \frac{e^{-2 \ome}}{2
F} V. 
\eeq
Then, performing on $\f$ the transformation given by (\ref{14}) and on 
$V$ the transformation
\beq
\label{19b}
W(\fbr(\f))= -\frac{e^{-2 \ome(F)}}{2 F} V
\eeq
in which $\ome(F)$ satisfies (\ref{17b}), Eq. (\ref{18b}) becomes
\beq
\label{20b}
{\bar{G}}\dmunu= \fbr\ddemu \fbr\ddenu- \half \gbrd 
\fbr\ddea \fbr\udea- \gbrd W,
\eeq
which correspond to the Einstein equations in presence of a scalar 
field $\fbr$ with potential $W$. The expression for $\ome(F)$ is 
easily obtained from (\ref{17b}), that is 
\beq
\label{21b}
\ome= \half \ln{F}+ \ome_0
\eeq
in which $\ome_0$ is the integration constant. The potential $W$ takes 
the form
\beq
\label{22b}
W= -\frac{V}{2 \al F}.
\eeq
Comparing (\ref{22b}) with the second of (\ref{14}), we see that, fixing
$\al= -2$, the definition of $W$ coincides with that one of $\VBR$. We
have then the full compatibility with the Lagrangian approach
obtaining for $\ome$ the same relation as (\ref{12}); in this sense we
have verified the full equivalence between the NMC and the
Einstein--Hilbert Lagrangian density plus scalar field.

Our final remark regards the relations (\ref{14}): actually, from
(\ref{13}) the relation between $\fbr\ddea$ and $\f\ddea$ present a
$\pm$ sign in front of the square root, which corresponds to have the
same or opposite sign in the derivative of $\f$ and $\fbr$ with
respect to $x\da$. What follows is independent of such sign; we will
choose then the positive one. 

\section{\normalsize \bf The cosmological case}

Let us assume now that the spacetime manifold is described by a FRW
metric, that is we consider homogeneous and isotropic cosmology. Then
the Ricci scalar, has the expression $R= -6 \disp{ \left(
\frac{\add}{a}+ \frac{\ad^2}{a^2}+ \frac{\ka}{a^2}\right)}$, in which
the dot means the derivative with respect to time and $\ka$ is the
curvature constant. The Lagrangian density (\ref{1}) takes the form
\bib{nuovocim} 
\beq
\label{17}
L_t= 6 F(\f) a \ad^2+ 6 \FDEF(\f) a^2 \ad \fd- 6 F(\f) a \ka+ \half 
a^3 \fd^2- a^3 V(\f).
\eeq
Expression (\ref{17}) can be seen as a ``point--like'' Lagrangian on
the configuration space $(a,~\f)$ (in this way the meaning of the
expression we used in the introduction is clarified). With the
subscript $t$, we mean that the time--coordinate considered is the
universal time $t$: this remark is important for the forthcoming
discussion. The Euler--Lagrange equations relative to (\ref{17}) are
then 
\beq
\label{18}
\left \{
\begin{array}{l}
\disp{ \frac{2 \add}{a}+ \frac{\ad^2}{a^2}+ \frac{2 \FDEF \ad \fd}{F
a}+ \frac{\FDEF \fdd}{F}+ \frac{\ka}{a^2}+ \frac{F_{\f \f} \fd^2}{F}-
\frac{\fd^2}{4 F}+ \frac{V}{2 F}= 0} \\ 
~ \\
\disp{ \fdd+ \frac{3 \ad \fd}{a}+ \frac{6 \FDEF \ad^2}{a^2}+ \frac{6
\FDEF \add}{a}+ \frac{6 \FDEF \ka}{a^2}+ \VDEF= 0} 
\end{array}
\right.
\eeq
which correspond to the (generalized) second order Einstein equation and
the Klein--Gordon equation in the FRW case. The energy function 
relative to (\ref{17}) is 
\beq
\label{19}
\begin{array}{ll}
E_t & \disp{ =\frac{\pa L_t}{\pa \ad} \ad+ \frac{\pa L_t}{\pa \fd} \fd- 
L_t=} \\
~ & ~ \\
~ & =6 F a \ad^2+ 6 \FDEF a^2 \ad \fd+ 6 F a \ka+ \half a^3 \fd^2+ a^3
V,
\end{array}
\eeq
and we see that the first order generalized Einstein equation is
equivalent to 
\beq
\label{20}
E_t= 0.
\eeq
Performing the conformal transformation defined by (\ref{10}),
(\ref{12}), (\ref{14}) on the FRW metric, one should obtain the
corresponding expression for the Lagrangian and the corresponding
equations of the Einstein--cosmology from the NMC Lagrangian
(\ref{17}) and from the generalized Einstein and Klein--Gordon
equations, respectively. Unfortunately we see that the presence of the
conformal factor (\ref{12}) implies that the transformed line element
which is obtained is no longer expressed in the ``universal time
form''. Actually the scale factor of the Einstein theory can be
defined as the scale factor of the NMC theory multiplied by the
conformal factor, but the time coordinate of the Einstein theory has
to be redefined if we require to have an ``universal time''. Absorbing
the conformal factor in the redefinition of time, we obtain the
transformation on the time coordinate. Therefore, the transformation
from the Jordan frame to the Einstein frame in the cosmological case
is given by 
\beq
\label{21}
\left \{
\begin{array}{l}
\abr= \sqrt{-2 F(\f)} a \\
~ \\
\disp{ \frac{d\fbr}{dt}= \sqrt{\frac{3 \FDEF^2- F}{2 F^2}}\, 
\frac{d\f}{dt}} \\ 
~ \\
d\tbr= \sqrt{-2 F(\f)}\, dt.
\end{array}
\right.
\eeq
From the first and the third of (\ref{21}) we have that, on the
Jordan--frame solutions $a(t)$, $\f(t)$, we obtain $\abr$ as a
function of $\tbr$ only; indeed the important thing is the fact
that the equations for $\bar{a}$ we will obtain are the standard
Einstein equations. The second of (\ref{21}) corresponds to relation
(\ref{14}) under the given assumption of homogeneity and isotropy.
Under transformation (\ref{21}) we have that 
\beq
\label{22}
\begin{array}{ll}
\disp{ \frac{1}{\sqrt{-2 F}} L_t} & \disp{ =\frac{1}{\sqrt{-2 F}}}
\left( 6 F a \ad^2+ 6 \FDEF a^2 \ad \fd- 6 F a \ka+ \half a^3 \fd^2-
a^3 V\right)=
\\ 
~ & ~ \\
~ & =-3 \abr \abrd^2+ 3 \ka \abr+ \half \abr^3 \fbrd^2- \abr^3
\VBR(\fbr)= \LBR_\tbr 
\end{array}
\eeq
in which the dot over barred quantities means the derivative with
respect to $\tbr$; $L_t$ is given by (\ref{17}) and $\LBR_\tbr$
coincides with the ``point--like'' Lagrangian obtained from the
Einstein--Hilbert action plus a scalar field under the assumption of
homogeneity and isotropy. In this way the invariance of the homogeneus
and isotropic action under (\ref{21}) is insured, being $L_t$ and
$\LBR_\tbr$ connected by the (\ref{22}). The same correspondence as
(\ref{22}) exists between the energy function $E_t$ and $\EBR_\tbr$,
that is, there is correspondence between the two first order Einstein
equations in the two frames. We focus now our attention on the way the
Euler--Lagrange equations transform under (\ref{21}). The
Euler--Lagrange equations relative to (\ref{22}) are the usual second
order Einstein equation and Klein--Gordon equation 
\beq
\label{23}
\left \{
\begin{array}{l}
\disp{ \frac{2 \abrdd}{\abr}+ \frac{\abrd^2}{\abr}+
\frac{\ka}{\abr^2}+ \half \fbrd^2- \VBR= 0} \\ 
~ \\
\disp{ \fbrdd+ \frac{3 \abrd \fbrd}{\abr}+ \VBRDEF= 0}.
\end{array}
\right.
\eeq
Under (\ref{21}) it is straighforward to verify that they become
\beq
\label{24}
\left \{
\begin{array}{l}
\disp{ \frac{2 \add}{a}+ \frac{\ad^2}{a^2}+ \frac{2 \FDEF \ad \fd}{F
a}+ \frac{\FDEF \fdd}{F}+ \frac{\ka}{a^2}+ \frac{F_{\f \f} \fd^2}{F}-
\frac{\fd^2}{4 F}+ \frac{V}{2 F}= 0} \\ 
~ \\
\disp{ \fdd+ \frac{3 \ad \fd}{a}+ \frac{6 \FDEF F_{\f \f}- \FDEF}{3
\FDEF^2- F} \frac{\fd^2}{2}+ \frac{2 \FDEF V}{3 \FDEF^2- F}- \frac{F
\VDEF}{3 \FDEF^2- F}= 0} 
\end{array}
\right.
\eeq
which do not coincide with the Euler--Lagrange equations given by
(\ref{18}). Using the first of (\ref{18}), the second of (\ref{24})
can be written as 
\beq
\label{25}
\begin{array}{l}
\disp{ \frac{F- 3 \FDEF^2}{F} \fdd+ \frac{3(F- 3 \FDEF^2)}{F}
\frac{\ad}{a} \fd+ \frac{\FDEF- 6 F_{\f \f} \FDEF}{F} \frac{\fd^2}{2}+
\frac{\FDEF \fd^2}{4 F}- \frac{2 \FDEF V}{F}+ \VDEF+} \\ 
~ \\
\disp{ +\frac{3 \FDEF \ad^2}{a^2}+ \frac{3 \FDEF \ka}{a^2}+ \frac{3
\FDEF^2 \ad \fd}{a}= 0} 
\end{array}
\eeq
which becomes, taking into account (\ref{19})
\beq
\label{26}
\begin{array}{l}
\disp{ \frac{F- 3 \FDEF^2}{F} \fdd+ \frac{3(F- 3 \FDEF^2)}{F}
\frac{\ad \fd}{a}+ \frac{1}{F} \frac{d}{d\f} (F- 3\FDEF^2)
\frac{\fd^2}{2}+ \frac{\FDEF \fd^2}{4 F}- \frac{2 \FDEF V}{F}+ \VDEF+}
\\ 
~ \\
\disp{ +\frac{\FDEF}{2 a^3 F} E_t= 0}. 
\end{array}
\eeq
Comparing (\ref{26}) with the second of (\ref{24}) we see that they
concide if $F- 3 \FDEF^2 \neq 0$ and $E_t= 0$. The quantity $F- 3
\FDEF^2$ is proportional to the Hessian determinant of $L_t$ with
respect to $(\ad,~\fd)$; we want this Hessian different from zero in
order to avoid pathologies in the dynamics \bib{cqg}, while $E_t= 0$
corresponds to the first order Einstein equation. It seems that, under
the assumption of homogeneity and isotropy and the request of having
the metric expressed in the universal time in both the Einstein and
Jordan frame, we have conformal equivalence between the
Euler--Lagrange Eqs. (\ref{18}) and (\ref{23}) only on the
(cosmological) solutions. Actually, if we look more carefully to this
problem, we notice that, making the hypotheses of homogeneity and
isotropy on the field Eqs. (\ref{2}) and (\ref{5}), we get the
generalized Einstein equations of first and second order, and the
Klein--Gordon equation. On the other side, the Euler--Lagrange
equations relative to (\ref{17}) are just the second order Einstein
equation and the Klein--Gordon equation, whereas the first order
Einstein equation is obtained from $E_t= 0$. Of course the same
happens in the Einstein frame. Therefore it is natural to expect that
the full conformal equivalence in the ``point--like'' formulation is
verified taking into consideration $E_t= 0$. 

It is possible to see more clearly at the problem of the cosmological
conformal equivalence, formulated in the context of the 
``point--like'' Lagrangian, if we use as time--coordinate the
conformal time $\eta$, connected to the universal time $t$ by the
usual relation 
\beq
\label{27}
a^2(\eta) d \eta^2= d t^2.
\eeq
We can see that the use of $\eta$ makes much easier the treatment of 
all the problems we have discussed till now.

The crucial point is the following: given the form of the FRW line
element expressed in conformal time $\eta$ one does not face the
problem of redefining time after performing a conformal
transformation, since in this case, the expansion parameter appears in
front of all the terms of the line element. From this point of view,
the conformal transformation which connects Einstein and Jordan frame
is given by 
\beq
\label{28}
\left \{
\begin{array}{l}
\abr= \sqrt{-2 F(\f)} a \\
~ \\
\disp{ \frac{d\fbr}{d\eta}= \sqrt{\frac{3 \FDEF^2- F}{2 F^2}}\, 
\frac{d\f}{d\eta}} 
\end{array}
\right.
\eeq
where $a$, $\f$, $\abr$, $\fbr$ are assumed as functions of $\eta$.

The Einstein--Hilbert ``point--like'' Lagrangian is given by
\beq
\label{29}
\LBR_\eta= -3 \abrpq+ 3 \ka \abr^2+ \half \abr^2 \fbrpq- \abr^4 
\VBR(\fbr)
\eeq
in which the prime means the derivative with respect to $\eta$, and 
the subscript $\eta$ means that the time--coordinate considered is the
conformal time. Under transformation (\ref{28}), it becomes 
\beq
\label{30}
\begin{array}{ll}
\LBR_\eta & =-3 \abrpq+ 3 \ka \abr^2+ \half \abr^2 \fbrpq- \abr^4 
\VBR(\fbr) \\
~ & ~ \\
~ & =6 F(\f) \ap+ 6 \FDEF(\f) a \ap \fp- 6 F(\f) \ka a^2+ \half a^2
\fpq- a^4 V(\f)= L_\eta 
\end{array}
\eeq
which corresponds to the ``point--like'' Lagrangian obtained from the
Lagrangian density in (\ref{1}) under the hypotheses of homogeneity
and isotropy, using the conformal time as time coordinate. 

This means that the Euler--Lagrange equations relative to (\ref{29}),
which coincides with the second order Einstein equation and the
Klein--Gordon equation in conformal time, correspond to the
Euler--Lagrange equations relative to (\ref{30}), under the
transformation (\ref{28}). Moreover, the energy function $\EBR_\eta$
relative to (29) corresponds to the energy function $E_\eta$ relative
to (\ref{30}), so that there is correspondence between the first order
Einstein equations. Furthermore, in order to have full coherence
between the two formulations, it is easy to verify that, both in the
Jordan frame and in the Einstein frame, the Euler--Lagrange equations,
written using the conformal time, correspond to the Euler--Lagrange
equations written using the universal time except for terms in the
energy function; for it one gets the relation 
\beq
\label{31}
E_\eta= a E_t
\eeq
which holds in both the frames; thus the first order Einstein equation
is preserved under the transformation from $\eta$ to $t$ and there is
full equivalence between the two formulations. We want to point out
that for the two Lagrangians $L_\eta$ and $L_t$ the same relation as
(\ref{31}) holds; this remark is useful for forthcoming
considerations.

\section{\normalsize \bf The presence of ordinary matter}

So far we have analysed the general conformal equivalence and the
cosmological conformal equivalence between Einstein frame and Jordan
frame in presence of a scalar field. What happens when ordinary matter
is present (see \bib{dicke})? We focus our attention on the
cosmological case. 

The standard Einstein (cosmological) ``point--like'' Lagrangian (when 
noninteracting scalar field and ordinary matter are present) is given 
by
\beq
\label{31b}
\LBR_{tot}= \LBR_\tbr+ \LBR_{mat}
\eeq
in which $\LBR_\tbr$ is given by (\ref{22}) and $\LBR_{mat}$ is the 
Lagrangian relative to matter. Using the contracted Bianchi identity,
it can be seen that $\LBR_{mat}$ can be written as \bib{matter} 
\beq
\label{32}
\LBR_{mat}= -D \abr^{3(1- \gam)}
\eeq
where $D$ is connected with the total amount of matter. In writing 
(\ref{31b}) and (\ref{32}) we have chosen the universal time as 
time--coordinate. Under the transformation (\ref{21}) we have, beside 
relation (\ref{22}), that (\ref{32}) corresponds to
\beq
\label{33}
\LBR_{mat}= (\sqrt{-2 F})^{3(1- \gam)} L_{mat}
\eeq
where, analogously to (\ref{32})
\beq
\label{34}
L_{mat}= D a^{3(1- \gam)}.
\eeq
Then we have that, under (\ref{21}), (\ref{31b}) becomes
\beq
\label{35}
\frac{1}{\sqrt{-2 F}} L_{tot}^{(1)} 
=\frac{1}{\sqrt{-2 F}} [L_t+ (\sqrt{-2 F})^{(4- 3 \gam)} L_{mat}] 
\eeq
in which we have defined the total ``point--like'' Lagrangian after
the conformal transformation as 
\beq
\label{36}
L_{tot}^{(1)}= L_t+ (\sqrt{-2 F})^{(4- 3 \gam)} 
L_{mat},
\eeq
(cfr. (\ref{22})); the transformation of $\LBR_{tot}$ under (\ref{21})
has to be written following the expression (\ref{35}) and consequently
the ``point--like'' Lagrangian $L_{tot}^{(1)}$, has to be defined as
in (\ref{36}). The use of the superscript $(1)$ for $L_{tot}$ will be
clarified in a moment. The factor $\frac{1}{\sqrt{-2 F}}$ in evidence
out of the square bracket, is introduced in order to preserve the
invariance of the reduced action under transformation (\ref{21}),
since that factor is also the one which appears in the
time--coordinate transformation in (\ref{21}). 

The Lagrangian (\ref{36}) could be then assumed to describe a
cosmological NMC--model with a scalar field and ordinary matter as
gravitational sources. By the way, we see that, unless $\gam=
\frac{4}{3}$, the standard matter Lagrangian term is coupled with the
scalar field in a way which depends on the coupling $F$. Such coupling
between the matter and the scalar field is an effect of the
transformation, therefore depending on the coupling $F$. This is one
way to look at the problem, but we can also proceed in a different way
to determine the Lagrangian in presence of matter. We can consider as
the total ``point--like'' Lagrangian 
\beq
\label{37}
L_{tot}^{(2)}= L_t+ L_{mat}.
\eeq
That is, we take the ``point--like'' Lagrangian of the NMC theories,
given by (\ref{17}), and add up to it the standard matter term defined
in (\ref{34}). It is clear now why we have introduced the notation
$L_{tot}^{(1)}$ and $L_{tot}^{(2)}$. Of course the full theory
described by (\ref{37}) is by no mean conformally equivalent to that
one described by (\ref{31b}). Also, the transformation does not give
rise to any coupling between matter and scalar field. One could just
point out that the matter term defined by (\ref{34}) has been obtained
in the context of the Einstein frame and in this sense it could be not
legitimate using it in a NMC theory. 

The problem of the physical system, also connected with the
formulation of the Equivalence Principle \bib{will}, has been already
discussed and it is well known, in particular in the case of the
higher order theories \bib{brans} \bib{sokolowski} \bib{francaviglia}
\bib{magnano-soko} (and references quoted therein), but also in the
context of NMC theories \bib{dicke} \bib{damour1}. In the case we are
considering, the problem still concerns the choice of the physical
system but from another point of view, since the Lagrangians
(\ref{36}) and (\ref{37}) are not connected by a conformal
transformation. The problem concerns which is the Lagrangian to
describe, in the Jordan frame, a cosmological model with a scalar
field plus ordinary matter, between the lagrangian $L_{tot}^{(1)}$ and
$L_{tot}^{(2)}$, once we assume that the physical system is that one
of a NMC theory (some authors consider the Jordan frame as the
physical one, see for examples \bib{jordan}, while in \bib{einstein}
the Einstein frame is the physical one). 

This sort of ambiguity can be clarified in the general context of the
field theory, introduced in Sec. 2. We focus our attention on the
contracted Bianchi identity. From the point of view of the field
equations, the choice of the Lagrangian $L_{tot}^{(2)}$ to describe
the gravitational field with a scalar field and ordinary matter (non
interacting with the scalar field) as sources in the Jordan frame
corresponds to write the field equations as 
\beq
\label{38b}
F(\f) G\dmunu= -\half T\dmunu- g\dmunu \Box_{\GAM} F(\f)+ F(\f)\ddemunu+
T_{(mat)~ \mu\nu}
\eeq
which is obtained just adding up the ordinary matter as a further
source term to to the field equations in the NMC case given by
(\ref{2}). Performing the covariant divergence of both sides and
taking into account of the expression of $T_{(\f)~ \mu\nu}$ given by
(\ref{4}), we get 
\beq 
\label{39b}
F\ddenu G\dmuunu+ \half \f\ddemu \f_{;~;\nu}^\mu+ \half \del\dmuunu 
\VDEF \f\ddenu+ \del\dmuunu (\Box_{\Gamma} F)\ddenu- 
F_{;\mu~\nu}^{~\nu~}= T_{(mat)~\mu~;\nu}^{~~~~~~~\nu}
\eeq
which can be written as
\beq
\label{40b}
F\ddenu R\dmuunu- \half R F\ddemu+ \half \f\ddemu (\Box_{\GAM} \f+
\VDEF)+ (\Box_{\Gamma} F)\ddemu- F_{;\mu~\nu}^{~\nu~}=
T_{(mat)~\mu~;\nu}^{~~~~~~~\nu} 
\eeq
where we have taken into account Eq. (\ref{3}). The last two
terms of the lefthand side of (\ref{40b}) give 
\beq
\label{41b}
\begin{array}{ll}
(g\uab F\ddeab)\ddemu- F_{;\mu~\nu}^{~\nu~} & =-g\uab R^\lam_{~\be\al\nu} 
F_{;\lam} \\
~ & ~ \\
~ & =-F_{;\lam} R_\mu^{~\lam}.
\end{array}
\eeq
Eq. (\ref{40b}), taking into account Eq. (\ref{41b}) gives
then an interesting relation, that is 
\beq 
\label{42b}
\half \f\ddemu (\Box_{\GAM} \f+ \VDEF- R \FDEF)=  
T_{(mat)~\mu~;\nu}^{~~~~~~~\nu}.
\eeq
Comparing relation (\ref{42b}) with Eq. (\ref{5}) we see that the 
lefthand side coincides with the lefthand side of the Klein--Gordon 
equation; it means that the continuity equation for the ordinary
matter holds, that is 
\beq
\label{43b}
T_{(mat)~\mu~;\nu}^{~~~~~~~\nu}= 0
\eeq
in which $T_{(mat)~\mu~;\nu}^{~~~~~~~\nu}$ is the energy--momentum 
tensor of the matter, relative to the NMC metric $\gd$. Thus, it means
that, choosing the Jordan frame as the physical frame and equations
(\ref{38b}) as field equations, the conservation of matter is relative
to the physical metric $\gd$. In this sense the legitimate way to
describe scalar field plus ordinary matter in the Jordan frame is the
one given by the field equations (\ref{38b}). The corresponding
action, in the cosmological case, corresponds to choose
$L_{tot}^{(2)}$ as Lagrangian \bib{brans} \bib{sokolowski}
\bib{francaviglia} \bib{magnano-soko} (and ref. quoted therein). 

We can only say that such considerations could be a hint for further
developments in the context of the Jordan frame (for a totally
different point of view see \bib{damour1}). 

The problem can be further analysed from the point of view of the
energy density parameter $\OME$. We can see in fact that the presence
of the coupling gives some contributions to $\OME$. Let us consider
the first order Einstein equation relative to the total lagrangian
$L_{tot}^{(2)}$ in the Jordan frame in presence of matter 
\beq
\label{44b}
\frac{\ad^2}{a^2}+ \frac{\FDEF \ad \fd}{F a}+ \frac{\ka}{a^2}+ \half 
\frac{\fd^2}{6 F}+ \frac{V}{6 F}+ \frac{D}{6 F a^{3 \gam}}= 0
\eeq
which can be seen as obtained from the standard first order Einstein
equation after the conformal transformation (\ref{21}), having just
added ordinary matter. The last term on the lefthand side being just
the effective energy density relative to matter; the factor
$-\frac{1}{2 F}$ represents the effective coupling. Taking into
account the definition of the Hubble parameter $H$, (\ref{44b}) can be
rewritten as 
\beq
\label{45b}
H^2+ \frac{\FD H}{F}+ \frac{1}{6 F} \left( \half \fd^2+ V\right)+ 
\frac{D}{6 F a^{3 \gam}}= 0
\eeq
(we consider the case $\ka= 0$). We get then 
\beq
\label{46b}
-\frac{\FD}{F H}- \frac{1}{6 H^2 F} \left( \half \fd^2+ V\right)- 
\frac{D}{6 H^2 F a^{3 \gam}}= 1
\eeq
after dividing by $H^2$. As usual, the righthand side is the total
energy density parameter $\OME_{tot}$ which is equal to 1, having
assumed $\ka= 0$. The last term on the lefthand side represents the
effective contribution to the density parameter due to the matter,
$\OME_{mat}= -\frac{D}{6 H^2 F a^{3 \gam}}$, while the term in
parentesis represents the effective energy density contribution due to
the scalar field, $\OME_{\f}= -\frac{1}{6 H^2 F} \left( \half \fd^2+
V\right)$; the first term is connected with the variation of the
coupling, $\OME_{coup}= -\frac{\FD}{F H}$. That is, there is a
contribution to the energy density parameter due to the presence of
the nonminimal coupling. This term, for what we have said, can be seen
as coming from the conformal transformation considered. The parameter
$\OME$ is an observative quantity (the present value of $\OME_{tot}$
is assumed to be equal to 1) thus, in principle, from its analysis one
could be able to infer whether the physical frame is the Jordan or the
Einstein frame (see also \bib{sokolowski}). 

A final remark we would like to do concerns the case $\gam=
\frac{4}{3}$. As we have said, performing the transformation
(\ref{21}) on (\ref{31b}) no coupling between the scalar field and the
matter is induced if the matter is a radiative perfect fluid: this
seems to be quite reasonable, since the particles which constitute a
radiative fluid have zero mass. 

\section{\normalsize \bf Conformal transformations and Noether 
symmetries} 

We want to analyse now the compatibility between the conformal
transformation we have considered so far and the presence of Noether
symmetries in the ``point--like'' Lagrangian in the configuration
space $(a,~\f)$, \ie in the cosmological case. Some of the authors, in
previous papers (see for examples \bib{cqg} \bib{mc}) have developed a
method to find exact cosmological solutions relative either to purely
scalar--tensor Lagrangians or to scalar--tensor Lagrangians with
ordinary matter, both in MC theories, when the Lagrangian is given by
(\ref{31b}), and in NMC theories, having taken (\ref{37}) as
Lagrangian. 

Now we want to analyse the problem whether the conformal
transformation connecting Einstein and Jordan frame preserves the
presence of a Noether symmetry. Since the existence of a Noether
symmetry implies the existence of a vector field $X$ along which
$\L_{X} L= 0$, this happens if the Lie derivative of the Lagrangian
along a vector field is preserved. We can see that the Lie derivative
is preserved under the conformal transformation considered, but only
in absence of ordinary matter. It turns out to be quite simple to be
verified if we choose as time--coordinate the conformal time. As we
have seen, we have that in absence of matter, using the time $\eta$
the ``point--like'' Lagrangian in the Einstein and Jordan frame given
by (\ref{29}) and (\ref{30}) respectively, correspond to each other
under the conformal transformation given by (\ref{28}). The second of
(\ref{28}), in principle, can be integrated, so that its finite form
together with the first of (\ref{28}) can represent a ``coordinate
transformation'' on the configuration space $(a,~\f)$. Thus, a given
lift--vector field of the form \bib{marmo} 
\beq
\label{38}
X_\eta= \al \frac{\pa}{\pa a}+ \be \frac{\pa}{\pa \f}+ \al^{'} 
\frac{\pa}{\pa \ap}+ \be^{'} \frac{\pa}{\pa \fp}
\eeq
in which $\al= \al(a,~\f)$, $\be= \be(a,~\f)$ corresponds under this 
transformation to the lift--vector field on the configuration space
$(\abr,~\fbr)$
\beq
\label{39}
\bar{X}_\eta= \bar{\al} \frac{\pa}{\pa \abr}+ \bar{\be}
\frac{\pa}{\pa \fbr}+ \bar{\al}^{'} \frac{\pa}{\pa \abrp}+
\bar{\be}^{'} \frac{\pa}{\pa \fbrp} 
\eeq
in which $\bar{\al}= \bar{\al}(\abr,~\fbr)$, $\bar{\be}= 
\bar{\be}(\abr,~\fbr)$ are connected to $\al= \al(a,~\f)$, $\be=
\be(a,~\f)$ through the Jacobian matrix relative to the ``coordinate
transformation'' defined by (\ref{28}). We remind that the prime means
the derivative with respect to the time $\eta$. The Lie derivative of 
$L_\eta$ along the vector field $X_\eta$ corresponds then to the Lie 
derivative of $\LBR_\eta$ along ${\bar{X}}_\eta$ \bib{tre-grazie}
\beq
\label{40}
\L_{X_\eta} L_\eta= \L_{\bar{X}_\eta} \LBR_\eta.
\eeq
Therefore, if $X_\eta$ is a Noether vector field relative to $L_\eta$ 
one has
\beq
\label{41}
\L_{X_\eta} L_\eta= 0
\eeq
and, from (\ref{40}), $\bar{X}_\eta$ is a Noether vector field 
relative to $\LBR_\eta$.

We have seen till now that the choice of $\eta$ as time--coordinate is 
convenient from a formal point of view, but, as we have already 
remarked, in order to analyse the phenomenology relative to a given 
model and to obtain then quantities comparable with the observational 
data, the appropriate choice of time--coordinate is the universal time 
$t$. The problem with the universal time is that it is not preserved 
by the conformal transformation, as we have pointed out in Sec. 2, 
thus the conformal transformation we consider does not take simply the 
form of a ``coordinate transformation'' on the phase space $(a,~\f)$, 
then its compatibility with the presence of a Noether symmetry cannot 
be easily verified. Of course it must hold also under such choice of 
time--coordinate.

We decide not to verify such compatibility directly. Rather, we
analyse how does the Lie derivative $\L_{X_\eta} L_\eta$ in the Jordan
frame is transformed under the time transformation (\ref{27}) which
connects $t$ with $\eta$. 

The explicit expression of $\L_{X_\eta} L_\eta$ is given by
\beq
\label{42}
\begin{array}{ll}
\L_{X_\eta} L_\eta & \disp{ =6 \left[ 2 F \frac{\pa \al}{\pa a}+
\left( \be+ a \frac{\pa \be}{\pa a}\right) \FDEF\right] \apq+ a
\left[ \al+ 6 \FDEF \frac{\pa \al}{\pa \f}+ a \frac{\pa \be}{\pa
\f}\right] \fpq+} \\ 
~ & ~ \\
~ & \disp{ +6 \left[ a \be F_{\f\f}+ \left( \al+ a \frac{\pa \al}{\pa
a}+ a \frac{\pa \be}{\pa \f}\right) \FDEF+ 2 F \frac{\pa \al}{\pa \f}+
\frac{a^2}{6} \frac{\pa \be}{\pa a}\right] \ap \fp+} \\ 
~ & ~ \\
~ & \disp{ -a^3 (4 \al V+ a \be \VDEF)- 6 a (2 F \al+ \FDEF a \be) 
\ka}
\end{array}
\eeq 
in which we have taken into account that
\beq
\label{43}
\al^{'}= \frac{\pa \al}{\pa a} \ap+ \frac{\pa \al}{\pa \f} \fp;~~~ 
\be^{'}= \frac{\pa \be}{\pa a} \ap+ \frac{\pa \be}{\pa \f} \fp;
\eeq
(\ref{42}) under the transformation (\ref{27}) becomes
\beq
\label{44}
\begin{array}{ll}
\L_{X_\eta} L_\eta & =\disp{ 6 \left[ 2 F \frac{\pa \al}{\pa a}+
\left( \be+ a \frac{\pa \be}{\pa a}\right) \FDEF\right] a^2 \ad^2+ a
\left[ \al+ 6 \FDEF \frac{\pa \al}{\pa \f}+ a \frac{\pa \be}{\pa
\f}\right] a^2 \fd^2+} \\ 
~ & ~ \\
~ & \disp{ +6 \left[ a \be F_{\f\f}+ \left( \al+ a \frac{\pa \al}{\pa
a}+ a \frac{\pa \be}{\pa \f}\right) \FDEF+ 2 F \frac{\pa \al}{\pa \f}+
\frac{a^2}{6} \frac{\pa \be}{\pa a}\right] a^2 \ad \fd+} \\
~ & ~ \\
~ & \disp{ -a^3 (4 \al V+ a \be \VDEF)- 6 a (2 F \al+ \FDEF a \be)
\ka} 
\end{array}
\eeq 
which can be written as
\beq
\label{45}
\begin{array}{ll}
\L_{X_\eta} L_\eta & \disp{ =6 a \left[ \al F+ 2 a F \frac{\pa
\al}{\pa a}+ a \left( \be+ a \frac{\pa \be}{\pa a}\right) \FDEF\right]
\ad^2+ a^3 \left[ \frac{3}{2} \al+ 6 \FDEF \frac{\pa \al}{\pa \f}+ a
\frac{\pa \be}{\pa \f}\right] \fd^2+} \\ 
~ & ~ \\
~ & \disp{ +6 a^2 \left[ a \be F_{\f\f}+ \left( 2 \al+ a \frac{\pa
\al}{\pa a}+ a \frac{\pa \be}{\pa \f}\right) \FDEF+ 2 F \frac{\pa
\al}{\pa \f}+ \frac{a^2}{6} \frac{\pa \be}{\pa a}\right] \ad \fd+} \\ 
~ & ~ \\
~ & \disp{ -a^3 (3 \al V+ a \be \VDEF)- 6 a (F \al+ \FDEF a \be) \ka+}
\\ 
~ & ~ \\
~ & \disp{ -6 \al F a \ad^2- \half \al a^3 \f^2- 6 \al \FDEF a^2 \ad
\fd- 6 \al F a \ka- \al a^3 V}. 
\end{array}
\eeq
The Lie derivative of $L_t$ given by (\ref{17}) along a lift--vector 
field of the form
\beq
\label{46}
X_t= \al \frac{\pa}{\pa a}+ \be \frac{\pa}{\pa \f}+ \dot{\al}
\frac{\pa}{\pa \ad}+ \bar{\be} \frac{\pa}{\pa \fd}
\eeq
is given by
\beq
\label{47}
\begin{array}{ll}
\L_{X_t} L_t & \disp{ =6 \left[ \al F+ 2 a F \frac{\pa \al}{\pa a}+ a
\left( \be+ a \frac{\pa \be}{\pa a}\right) \FDEF\right] \ad^2+ a^2
\left[ \frac{3}{2} \al+ 6 \FDEF \frac{\pa \al}{\pa \f}+ a \frac{\pa
\be}{\pa \f}\right] \fd^2+} \\ 
~ & ~ \\
~ & \disp{ +6 a \left[ a \be F_{\f\f}+ \left( 2 \al+ a \frac{\pa
\al}{\pa a}+ a \frac{\pa \be}{\pa \f}\right) \FDEF+ 2 F \frac{\pa
\al}{\pa \f}+ \frac{a^2}{6} \frac{\pa \be}{\pa a}\right] \ad \fd+} \\ 
~ & ~ \\
~ & \disp{ -a^2 (3 \al V+ a \be \VDEF)- 6 (F \al+ \FDEF a \be) \ka}
\end{array}
\eeq
in which we have taken into account that 
\beq
\label{48}
\dot{\al}= \frac{\pa \al}{\pa a} \ad+ \frac{\pa \al}{\pa \f} \fd;~~~ 
\dot{\be}= \frac{\pa \be}{\pa a} \ad+ \frac{\pa \be}{\pa \f} \fd.
\eeq
We remind that the dot means the derivative with respect to $t$. 

Comparing (\ref{45}) with (\ref{47}) and taking into account the 
expression of $E_t$ given by (\ref{19}) we obtain that, under the 
transformation (\ref{27}) the Lie derivative $\L_{X_\eta} L_\eta$ 
becomes
\beq
\label{49}
\L_{X_\eta} L_\eta= a \L_{X_t} L_t- (\L_{X_t} a) E_t,
\eeq
being $\L_{X_t} a= \al$.

It can be seen that the same relation as (\ref{49}) holds in the 
Einstein frame, that is 
\beq
\label{51}
\L_{\bar{X}_\eta} \LBR_\eta= \abr \L_{\bar{X}_\tbr} \LBR_\tbr- 
(\L_{\bar{X}_\tbr} \abr) \EBR_\tbr,
\eeq
with quite obvious meaning of $\bar{X}_\tbr$.

This implies that, if $X_\eta$ is a Noether vector field relative to
$L_\eta$, that is, if (\ref{41}) holds, the corresponding vector field
$X_t$ is such that 
\beq
\label{52}
\L_{X_t} L_t- \frac{\L_{X_t} a}{a} E_t= 0.
\eeq
It means also that, when the universal time is taken as
time--coordinate, the conformal transformation preserves the
expression given by the righthand side of (\ref{49}) and not the Lie
derivative along a given vector field $X_t$. Relation (\ref{52})
represents a more general way to express the presence of a first
integral for the Lagrangian $L_t$; associated to (\ref{52}) we have
the conserved quantity \bib{logan} 
\beq
\label{53}
-E_t \int \frac{\L_{X_t}}{a} dt+ \frac{\pa L_t}{\pa \ad} \al+ 
\frac{\pa L_t}{\pa \fd} \be= const 
\eeq
which, of course, holds on the solutions of the Euler--Lagrange 
equations. The vector field $X_t$ verifying (\ref{52}) can thus be 
seen as a generalized Noether vector field and the conformal 
transformation (\ref{21}) preserves this generalized symmetry. That 
is, if $X_t$ is a Noether vector field, in the sense of (\ref{52}), 
relative to $L_t$ then $\bar{X}_t$ is a Noether vector field relative 
to $\LBR_\tbr$ in the same sense, that is 
\beq
\label{54}
\L_{\bar{X}_\tbr} \LBR_\tbr- \frac{\L_{\bar{X}_t} \abr}{\abr} \EBR_t= 0.
\eeq
In terms of the conformal time, the first integral relative to (\ref{41}) 
for the Lagrangian $L_\eta$ is given by
\beq
\label{55}
\frac{\pa L_\eta}{\pa \ap} \al+ \frac{\pa L_\eta}{\pa \fp} \be=
const.
\eeq
We see that the expression (\ref{55}) corresponds to (\ref{53}) under 
the transformation (\ref{27}), except for a term in the energy 
function. In fact (\ref{55}) explicitely written is
\beq
\label{56}
(12 F \ap+ 6 \FDEF a \fp) \al+ (6 \FDEF a \ap+ a^2 \fp) \be= const,
\eeq
while (\ref{53}) is
\beq
\label{57}
-E_t \int \frac{\al}{a} dt+ (12 F a \ad+ 6 \FDEF a^2 \fd) \al+ (6 
\FDEF a^2 \ad+ a^3 \fd) \be= const.
\eeq
Taking into account (\ref{31}), we have that (\ref{57}), under 
(\ref{27}), becomes
\beq
\label{58}
-\frac{E_\eta}{a} \int \al d\eta+ (12 F \ap+ 6 \FDEF a \fp) \al+ (6 
\FDEF a \ap+ a \fp) \be= const
\eeq
which coincides with (\ref{56}) except for the term in $E_\eta$.

Therefore there is equivalence between the two formulations except for 
the term in $E_\eta$, coherently with what we have said at the end 
of Sec. 2.

As already said, some of the authors have formulated the existence of
a Noether vector field imposing 
\beq
\label{59}
\L_{X_t} L_t= 0
\eeq
using the universal time as time--coordinate; condition (\ref{59})
after the analysis we have done till now, turns out to be less general
than (\ref{54}). By the way, condition (\ref{59}) has the interesting
property that it implies the possibility to define some new
coordinates on the configuration space $(a,~\f)$, such that the
Lagrangian has a cyclic coordinate \bib{nuovocim} \bib{marmo},
reducing in this way the Euler--Lagrange equations. In fact, one can
always define new coordinate, say $(z,~w)$, in the configuration space
of the Lagrangian, such that the lift--vector field assumes the form
$X_t= \frac{\pa }{ \pa z}$, so that one has $\L_{X_t} L= \frac{\pa
L}{\pa z}$; in case (\ref{59}) holds one has then that $z$ is cyclic.
In the generalized case we are considering, it is no longer possible
to get this behavior, since $\L_{X_t} L_t \neq 0$ and consequently $z$
is no longer cyclic. In this case one has to use the first integral
(\ref{53}) together with the relation on the energy function to reduce
the Euler--Lagrange equations, that is, one has the system of Eqs.
(\ref{18}), Eq. (\ref{20}) and Eq. (\ref{53}). 

This problem corresponds in the Einstein frame to the system of 
equations (\ref{23}), the equation analogue to (\ref{20}), $\EBR_\tbr= 
0$, and the equation analogue to (\ref{53}),
\beq
\label{60}
-\EBR_\tbr \int \frac{\L_{\bar{X}_\tbr}}{\abr} d\tbr+ \frac{\pa
\LBR_\tbr}{\pa \abrd} \bar{\al}+ \frac{\pa \LBR_\tbr}{\pa \fbrd}
\bar{\be}= const. 
\eeq
Thus, finding the solutions of some cosmological model using the 
presence of a Noether symmetry (and therefore fixing the class of 
model compatible with it) in the Einstein frame, one gets via the 
conformal transformation (as given by (\ref{21})) the solutions to the
class of models in the Jordan frame corresponding to the one given in
the Einstein frame through the second of (\ref{14}). We are going to
give some significant examples in the following section. 

\section{\normalsize \bf Examples}

{\bf i)} Let us consider a quite easily solvable model in the
Einstein frame. We consider the cosmological model with a scalar
field, a constant potential and zero curvature. The Lagrangian is
given by 
\beq
\label{61}
\LBR_\tbr= -3 \abr \abrd^2+ \half \abr^3 \fbrd^2- \abr^3 
\LAM;
\eeq
the Euler--Lagrange equations and the zero energy function condition 
are given by
\beq
\label{62}
\left \{
\begin{array}{l}
\disp{ \frac{2 \abrdd}{\abr}+ \frac{\abrd^2}{\abr}+ \half \fbrd^2-
\LAM= 0} \\ 
~ \\
\disp{ \fbrdd+ \frac{3 \abrd \fbrd}{\abr}= 0}. 
\end{array}
\right.
\eeq
\beq
\label{63}
\frac{\abrd^2}{\abr^2}- \frac{1}{3} \left( \half \fbrd^2+ \LAM\right)= 0.
\eeq
The solutions are (see also \bib{marek})
\beq
\label{70}
\left \{
\begin{array}{l}
\abr= \left[ c_1 e^{\sqrt{3 \LAM} \, \tbr}- \disp{ \frac{\fbrd_0^2}{8
\LAM c_1^2}} e^{-\sqrt{3 \LAM} \, \tbr}\right]^{\frac{1}{3}} \\ 
~ \\
\fbr= \fbr_0+ \disp{ \sqrt{\frac{2}{3}}}\, \ln{\frac{1- \disp{
\frac{\fbrd_0}{2 c_1 \sqrt{2 \LAM}}} e^{-\sqrt{3 \LAM} \, \tbr}}{1+
\disp{ \frac{\fbrd_0}{2 c_1 \sqrt{2 \LAM}}} e^{-\sqrt{3 \LAM} \,
\tbr}}} 
\end{array}
\right.
\eeq
Of course only three constants of integration appear in the solution,
since Eq. (\ref{63}) corresponds to a contraint on the value of
the first integral $\EBR_\tbr$. We have that, in the limit of $\tbr
\rightarrow +\infty$, the behavior of $\abr$ is exponential with
characteristic time given by $\sqrt{\frac{\LAM}{3}}$, as we would
expect (see also \bib{lambda}), and $\fbr$ goes to a constant. 

Looking at the second of (\ref{14}) we have that such a model in the
Einstein frame corresponds in the, Jordan frame, to the class of
models with (arbitrarly given) coupling $F$ and potential $V$
connected by the relation 
\beq
\label{71}
\frac{V}{4 F^2}= \LAM,
\eeq
the solution of which can be obtained from (\ref{70}) via the 
transformation (\ref{21}). We can thus fix the potential $V$ and 
obtain from (\ref{71}) the corresponding coupling. This can be used as 
a method to find the solutions of NMC models with given potentials, 
the coupling being determined by (\ref{71}). We consider, as an 
example, the case 
\beq
\label{72}
V= \lam \f^4,~~~ \lam> 0
\eeq 
which correspond to a ``chaotic inflationary'' potential \bib{infl}. The
corresponding coupling is quadratic in $\f$ 
\beq
\label{73}
F= k_0 \f^2
\eeq
in which
\beq
\label{74}
k_0= -\half \sqrt{\frac{\lam}{\LAM}}
\eeq
Substituting (\ref{73}) into (\ref{21}) we get 
\beq
\label{75}
\left \{
\begin{array}{l}
\disp{ a= \frac{\abr}{\f \sqrt{-2 k_0}}} \\
~ \\
\disp{ d\f= \f \sqrt{\frac{2 k_0}{12 k_0 -1}}\, d\f} \\ 
~ \\
\disp{ dt= \frac{d\tbr}{\f \sqrt{-2 k_0}}}.
\end{array}
\right.
\eeq
As we see from these relations, it has to be $k_0< 0$. Integrating the
second of (\ref{75}) we have $\f$ in terms of $\fbr$ 
\beq
\label{76}
\f= \al_0 e^{\sqrt{\frac{2 k_0}{12 k_0- 1}} \, \fbr}.
\eeq
Substituting (\ref{76}) in the first of (\ref{75}) and taking into 
account the second of (\ref{70}) we have the solutions $a$ and $\f$ 
as functions of $\tbr$
\beq
\label{78}
\left \{
\begin{array}{l}
\f= \f_0 \left[ \frac{1- \disp{ \frac{\fbrd_0}{2 c_1 \sqrt{2 \LAM}}}
e^{-\sqrt{3 \LAM} \, \tbr}}{1+ \disp{ \frac{\fbrd_0}{2 c_1 \sqrt{2
\LAM}}} e^{-\sqrt{3 \LAM} \, \tbr}}\right]^{\sqrt{\frac{4 k_0}{3(12 k_0-
1)}}} \\ 
~ \\
a= \disp{ \frac{1}{\f_0 \sqrt{-2 k_0}}} \left[ c_1 e^{\sqrt{3 \LAM}
\, \tbr}- \disp{ \frac{\fbrd_0^2}{8 \LAM c_1^2}} e^{-\sqrt{3 \LAM}
\, \tbr}\right]^{\frac{1}{3}} \left[ \frac{\disp{ 1+ \frac{\fbrd_0}{2 c_1
\sqrt{2 \LAM}}} e^{-\sqrt{3 \LAM} \, \tbr}}{\disp{ 1- \frac{\fbrd_0}{2
c_1 \sqrt{2 \LAM}}} e^{-\sqrt{3 \LAM} \, \tbr}}\right]^{\sqrt{\frac{4
k_0}{3(12 k_0- 1)}}} 
\end{array} 
\right.
\eeq
in which $\f_0= \al_0 e^{\sqrt{\frac{2 k_0}{12 k_0- 1}} \, \fbr_0}$.
Substituting (\ref{76}) in the third of (\ref{75}), taking into 
account of (\ref{70}), we get
\beq
\label{79}
dt= \frac{d\tbr}{\f_0 \sqrt{-2 k_0}} \left[ \frac{1+ \disp{
\frac{\fbrd_0}{2 c_1 \sqrt{2 \LAM}}} e^{-\sqrt{3 \LAM} \, \tbr}}{1-
\disp{ \frac{\fbrd_0}{2 c_1 \sqrt{2 \LAM}}} e^{-\sqrt{3 \LAM}
\, \tbr}}\right]^{\sqrt{\frac{4 k_0}{3(12 k_0- 1)}}}. 
\eeq
We can obtain $\tbr$ as a function of $t$ integrating (\ref{79}) and 
then considering the inverse function; (\ref{79}) could be easily 
integrated if the exponent $\sqrt{\frac{4 k_0}{3(12 k_0- 1)}}$ 
would be equal to $\pm 1$, but this corresponds to a value of $k_0= 
\frac{3}{32}$ which is positive and thus it turns out to be not 
acceptable. In general, (\ref{79}) is not of easy solution. We can 
analyse its asymptotic behavior, obtaining
\beq
\label{80}
\frac{dt}{d\tbr} \stackrel{\tbr \rightarrow +\infty}{\rightarrow} 
\frac{1}{\f_0 \sqrt{-2 k_0}} 
\eeq
that is, asymptotically
\beq
\label{81}
t- t_0 \simeq \frac{\tbr}{\f_0 \sqrt{-2 k_0}}.
\eeq
Substituting (\ref{81}) in the asymptotic expression of (\ref{78}), we
obtain the asymptotic behavior of the solutions (since from (\ref{80})
one has $t \stackrel{\tbr \rightarrow +\infty}{\rightarrow} +\infty$) 
\beq
\label{83}
\left \{
\begin{array}{l}
a \simeq \disp{ \frac{c_1^{1/3}}{\f_0 \sqrt{-2 k_0}}} e^{\f_0
\sqrt{\frac{-2 
\LAM k_0}{3}} \, (t- t_0)} \\ 
~ \\
\f \simeq \f_0.
\end{array}
\right.
\eeq
Thus we have that, asymptotically, $a(t)$ is exponential as it had to be
(cfr. \bib{lambda}), and $\f(t)$ is constant; the coupling $F$ is
asymptotically constant too, so that, fixing the arbitrary constant of
integration to obtain the finite transformation of $\abr$, $\fbr$
(that is, fixing the units, see \bib{sokolowski}), once $k_0$ is fixed,
it is possible to recover asymptotically the Einstein gravity. 

As a remark we would like to notice that the asymptotic expression
(\ref{83}) of $a(t)$ and $\f(t)$ are solutions of the Einstein
equations and Klein--Gordon equation with zero curvature and $F$ given
by (\ref{72}), (\ref{73}). They have not been obtained as solutions of
the asymptotic limits of these equations. It means then that they are,
in any case, particular solutions of the given NMC--model. 

{\bf ii)} Another interesting case is the Ginzburg--Landau potential 
\beq
\label{84}
V= \lam (\f^2- \mu^2)^2,~~~ \lam> 0.
\eeq
The corresponding coupling is given by
\beq
\label{85}
F= k_0 (\f^2- \mu^2)
\eeq
in which $k_0$ is given by (\ref{74}) when $\f^2> \mu^2$ while is
given by (\ref{74}) with opposite sign when $\f^2< \mu^2$, in order to
have $F< 0$. With this coupling the corresponding conformal
transformation turns out to be singular for $\f^2= \mu^2$, thus with
this method it is not possible to solve this model for $\f$ equal to
the Ginzburg--Landau mass $\mu$. 

In this case, it is not so straightforward to get the explicit function
$\f= \f(\fbr)$ as in the previous case, since one should perform and
then obtain the inverse of the integral 
\beq
\label{86}
\fbr- \fbr_0= \int \frac{[3 \sqrt{\frac{\lam}{\LAM}}\f^2+ \half
(\f^2- \mu^2)]^\half}{[\frac{\lam}{4 \LAM}]^{\frac{1}{4}} (\f^2-
\mu^2)}\, d\f. 
\eeq
It is not so difficult to integrate (\ref{86}) \bib{grad}, the
difficulty raises in finding the inverse, which is needed to obtain
$t$ in terms of $\tbr$. By the way, analysing the integrand of
(\ref{86}), that is $\frac{d\fbr}{d\f}$, we can say that, except for
$\f^2= \mu^2$, the function $\fbr= \fbr(\f)$ is invertible; in
particular, since asymptotically $\fbr$ is constant, so is $\f$. Thus
it is possible to carry out a reasoning analogous to the previous one,
concluding that asymptotically the behavior of $a(t)$ is exponential
and that of $\f(t)$ is constant. 

{\bf iii)} We want to consider now the $V= \lam \f^4$ case from the
point of view of the Noether symmetries. We want to show that, in the
context of generalized Noether symmetries the NMC--model with
quartic potential and negative quadratic coupling admits a Noether
symmetry, while such a result has been not found in the previous
analysis of Noether symmetries (see \bib{nuovocim} \bib{cqg}). 

We have seen that the corresponding case in the Einstein frame is that
one of constant potential. The system of equations for the Noether
vector field obtained from (\ref{54}) is given by 
\beq
\label{87}
\left \{
\begin{array}{l}
\disp{ \frac{\pa \bar{\al}}{\pa \abr}= 0} \\
~ \\
\disp{ \bar{\al}+ \abr \frac{\pa \bar{\be}}{\pa \fbr}= 0} \\
\disp{ 6 \frac{\pa \bar{\al}}{\pa \fbr}- \abr^2 \frac{\pa \bar{\be}}{\pa 
\abr}=0} \\
~ \\
4 \al \VBR+ \abr \bar{\be} \VBRDEF= 0.
\end{array}
\right.
\eeq
Substituting $\VBR= \LAM$ in the fourth of (\ref{87}) one gets
$\bar{\al}= 0$; from the second one gets $\bar{\be}= const$; the first
and the third turn out to be identically verified. It is immediate to
see that the Lagrangian (\ref{61}) presents a Noether symmetry, since
it does not depend on $\fbr$; being, in this particular case,
$\L_{\bar{X}_\tbr} \abr= \bar{\al}= 0$, this is compatible, for what
we have already said, with the presence of a cyclic coordinate in the
Lagrangian. Performing the conformal transformation given by
(\ref{75}) on the Noether vector field 
\beq
\label{88}
\left \{
\begin{array}{l}
\bar{\al}= 0 \\
~ \\
\bar{\be}= \bar{\be}_0. 
\end{array}
\right.
\eeq
where $\bar{\be}_0$ is arbitrary, we obtain
\beq
\label{89}
\left \{
\begin{array}{l}
\disp{ \al= -a \sqrt{\frac{2 k_0}{12 k_0- 1}}\, \bar{\be}_0} \\
~ \\
\disp{ \be= \f \sqrt{\frac{2 k_0}{12 k_0- 1}}\, \bar{\be}_0}.
\end{array}
\right.
\eeq
As it has been shown in the previous section, (\ref{89}) is a Noether
vector field relative to the corresponding Lagrangian in the Jordan
frame, with potential given by (\ref{72}) and coupling given by
(\ref{73}). It is easy to verify that (\ref{52}) holds. 

{\bf iv)} There is another interesting case we would like to
quote in the context of inflationary models, as last example, \ie 
\beq
\label{90}
V= \lam \f^2,~~ \lam> 0;~~~ F= k_0 \f^2,~~ k_0< 0
\eeq
in the Jordan frame. Since the coupling is the same as the previous
case (cfr. (\ref{73})), the relative conformal transformation is given
by (\ref{75}). To obtain the corresponding potential in the Einstein
frame we have to substitute (\ref{76}) in the relation 
\beq
\label{91}
\VBR(\fbr)= \frac{\lam}{4 k_0^2 \f^2(\fbr)}
\eeq
that is
\beq
\label{92}
\VBR(\fbr)= \frac{\lam}{4 k_0^2 \f^2_0} e^{-2 \sqrt{\frac{2 k_0}{12
k_0- 1}} \, \fbr}. 
\eeq
We see that this case corresponds in the Einstein frame to the case of
the exponential potential, for which a particular solution is the
power--law inflation \bib{lucchin-mat} \bib{barrow}. The choice of the
sign that Lucchin and Matarrese do in \bib{lucchin-mat} corresponds to
the choice of the sign we have done at the end of Sec. 1. 

Finally, we would like to make a general remark connected with all the
examples we have treated, concerning the relation between the Hubble
parameter in the Einstein and in the Jordan frame. Such relation is
given by 
\beq
\label{32b}
\begin{array}{ll}
\bar{H}= \disp{ \frac{\abrd}{\abr}}= & ~ \\ 
~ & ~ \\
\disp{ =\frac{1}{(-2 F)} \left( -\frac{\FD}{\sqrt{-2 F}}+ \sqrt{-2 F}
\frac{\ad}{a}\right)} & \disp{ =\frac{\FD}{2 F \sqrt{-2 F}}+
\frac{H}{\sqrt{-2 F}}} 
\end{array}
\eeq
in which we have used (\ref{21}) and the definition of the Hubble
parameter. Relation (\ref{32b}) is quite useful to make some
considerations on the asymptotic behavior of the Hubble parameter (see
also \bib{lambda}): for examples, if we require an asymptotic de
Sitter--behavior in both the Einstein and Jordan frame, that is, we
require $\bar{H} \stackrel{\tbr \rightarrow +\infty}{\rightarrow}
\bar{C}$ and $H \stackrel{\tbr \rightarrow +\infty}{\rightarrow} C$
where $\bar{C}$ and $C$ are constants, from (\ref{32b}) we obtain a
differential equation for the coupling $F$ as a function of $t$ ($t>> 
0$), given by 
\beq
\label{33b}
\FD+ 2 C F- 2 \bar{C} F \sqrt{-2 F}= 0.
\eeq
Its solution is 
\beq
\label{34b}
F= -\frac{C^2}{2 \bar{C}^2} \left[ \frac{1}{1- F_0 e^{C\, t}}-
1\right]^2 
\eeq
in which $F_0$ is the integration constant; this is the time--behavior
that $F$ has to assume on the solution $\f(t)$, in order to have a de
Sitter asymptotical behavior in both frames. We see from (\ref{34b}) 
that, asymptotically, we recover the standard gravity.

\section{\normalsize \bf  Conclusions}

We have analysed the conformal equivalence between Jordan frame and
Einstein frame for general coupling functions and potentials, and we
have seen that any NMC theory assumes the form of the Einstein theory
with a scalar field as source of the gravitational field, provided the
metric undergoes the conformal transformation defined by (\ref{12})
and the scalar field and the potential are transformed according to
(\ref{14}) We see, from these transformations, that the scalar field,
although being scalar with respect to the coordinate tranformations of
the space--time by definition, is not conformally scalar. 

We have considered such equivalence more carefully in the cosmological
case, and we have seen that the conformal transformation in this case
takes the form given by (\ref{21}), in which also the time--coordinate 
is transformed. The transformation of the time--coordinate turns out 
to be necessary if one requires the time--coordinate in the Einstein 
frame to be the universal time as well. We have seen that in case one
chooses the conformal time as time--coordinate the transformation
defined by (\ref{12}) reduces to the form (\ref{28}), which can be
seen as a ``coordinate transformation'' on the configuration space
$(a,~\f)$; with such a choice of time--coordinate the conformal
equivalence between Jordan and Einstein frame in the cosmological case
turns out to be very simple to verify. 

The situation changes when ordinary matter is considered besides the
scalar field: the conformal equivalence in this case is broken. We
have analysed the possible descriptions of NMC theories in presence of
ordinary matter and we have seen how (\ref{43b}) could be taken as
hint in the definition of the appropriate Lagrangian, in agreement
with the current discussions about the compatibility of NMC theories
and the different formulations of the Equivalence Principle. 

We have also seen that if a Noether symmetry is present in the
``point--like '' cosmological Lagrangian, this is preserved by the
conformal transformation which connects Jordan and Einstein frames.
This has been formally formulated in the conformal time through
relation (\ref{41}), since in this case the problem of the redefining
the time--coordinate does not exist. We have then analysed the problem
in the universal time and we have seen that the Noether symmetry is
preserved under the generalized form (\ref{52}), which implies as well
the presence of a first integral for the corresponding Lagrangian. 

We have thus analysed some aspects of the conformal equivalence 
between Jordan frame and Einstein frame, in particular in the 
cosmological case. Moreover we have generalized and improved a method 
of solution of cosmological NMC--models, having shown 
that the conformal transformation considered preserves a Noether 
symmetry present in the ``point--like'' Lagrangian, in the sense of 
(\ref{52}). The forthcoming steps will be to investigate more deeply
the implications of that: it would be interesting to classify the 
classes of NMC theories which are solvable by this method, and also 
to understand if it is possible to characterize the inflationary 
solutions in such context, on one side; to apply this method and to
analyse the phenomenology of the models with couplings and potentials 
of physical interest, which can be solved, on the other side. This can
be done also in connection with the problem of an opportune
redefinition of the ``cosmological constant'', which in this context
can be time--dependent (\bib{lambda} see also \bib{moffat}). 

\vspace{.3truecm}
\begin{centerline}
{\bf AKNOWLEDGMENTS}
\end{centerline}
We would like to thank Marmo G for the useful discussions.

\vspace{.3truecm}

\begin{centerline}
{\bf REFERENCES}
\end{centerline}
\begin{enumerate}
\item\label{brans-dicke}
Brans C and Dicke R H 1961 \pr {\bf 124} 925
\item\label{jordan-fierz}
Jordan P 1955 {\it Schwerkraft und Weltall} (Braunschweig: Vieweg); 
1959 {\bf 157} 112 \\
Fierz M 1956 \hpa {\bf 29} 128
\item\label{dicke}
Dicke R H 1962 \pr {\bf 125} 2163
\item\label{nmc}
Bekenstein J D 1974 \aph {\bf 82} 535 \\
Singh T and Singh Tarkeshwar 1987 \ijmp {\bf A 2} 645 \\
Schmidt H J 1988 \pl {\bf 214B} 519 \\
Madsen M S 1988 \cqg {\bf 5} 627 \\
Damour T and Esposito-Far\`ese G 1992 \cqg {\bf 9} 2093 \\
\item\label{damour1}
Damour T, Gibbons G W and Gundlach C 1990 \prl {\bf 64} 123
\item\label{damour2}
Damour T and Nordtvedt K 1993 \pr {\bf D 48} 3436
\item\label{qnmc}
Penrose R 1965 \prs {\bf A 284} 159
Callan C G, Coleman S and Jackiw R 1970 \aph {\bf 59} 42 \\
Eastwood M and Singer M 1985 \pl {\bf 107A} 73 \\
Sonego S and Faraoni V 1993 \cqg {\bf 10} 1185
\item\label{ordsup}
Higgs P W 1959 \ncim {\bf 11} 816 \\
Whitt B 1984 \pl {\bf 145B} 176 \\
Ferraris M, Francaviglia M and Magnano G 1988 \cqg {\bf 5} L95 \\
Magnano G, Ferraris M and Francaviglia M 1987 \grg {\bf 19} 465 \\
Jakubiec A and Kijowski J 1988 \pr {\bf B 37} 1406
\item\label{qordsup}
Starobinskii A A 1980 \pl {\bf 91 B} 99 \\
Barrow J and Ottewill A C 1983 {\it J. Phys. A: Math. Gen} {\bf 16} 
2757 \\
Miji\'c M B, Morris M S and Suen W M 1986 \pr {\bf D 34} 2934 \\
\item\label{birrel}
Birrel N D and Davies P C W 1982 {\it Quantum Fields in Curved 
Spacetimes} (Cambridge: Cambridge University Press) \\
\item\label{brans}
Brans C H 1988 \cqg {\bf 5} L197 
\item\label{sokolowski}
Sokolowski L M 1989 \cqg {\bf 6} 2045
\item\label{francaviglia}
Ferraris M, Francaviglia M and Magnano G 1990 \cqg {\bf 7} 261
\item\label{magnano-soko}
Magnano G and Sokolowski L M 1994 \pr {\bf D 50} 5039
\item\label{nuovocim}
Capozziello S, de Ritis R, Rubano C and Scudellaro P 1996 La Riv. del 
Nuovo Cim. {\bf 4} 
\item \label{wald}
Synge J L 1960 {Relativity. The general theory} (Amsterdam: North 
Holland) \\
Parker L 1973 \pr {\bf D 7} 976 \\
Wald R M 1984 {\it General Relativity} (Chicago: University of Chicago 
Press)
\item\label{cqg}
Capozziello S and de Ritis R 1993 \pl {\bf 177A} 1 \\
Capozziello S and de Ritis R 1994 \cqg {\bf 11} 107
\item\label{lambda}
Capozziello S and de Ritis R 1996 astro-ph 9605070 \\
Capozziello S, de Ritis R and Marino A A 1996 \hpa {\bf 69} 241
\item\label{matter}
Capozziello S and de Ritis R 1994 \pl {\bf 195A} 48 
\item\label{will}
Will C M 1993 {\it Theory and experiments in gravitational physics},
(Cambridge: Cambridge University Press) 
\item\label{jordan}
Bergmann P G 1968 \ijtp {\bf 1} 25 \\
Piccinelli G, Lucchin F and Matarrese S 1992 \pl {\bf 277B} 58 \\
Mollerach S and Matarrese S 1992 \pr {\bf D 45} 1961 \\
Barrow J D and Maeda K 1990 \np {\bf B 341} 294 \\ 
Barrow J D 1993 \pr {D 47} 5329 \\
Cotsakis S and Flessas G 1993 \pr {\bf D 48} 3577
\item\label{einstein}
Salopek D, Bond J and Bardeen J 1989 {\bf D 40} 1753 \\
Gibbons G W and Maeda K 1988 \np {\bf B 298} 741 \\
Kolb E W, Salopek D and Turner M 1990 \pr {D 412} 3925 \\
Deruelle N, Garriga J and Verdaguer E 1991 \pr {\bf D 43} 1032
\item\label{mc}
de Ritis R, Marmo G, Platania G, Rubano C, Scudellaro P and  Stornaiolo 
C 1990 \pr {\bf D 42} 1091
\item\label{marmo}
Abraham R and Marsden J 1978 {\it Foundation of Mechanics} (New York: 
Benjamin) \\
Marmo G, Saletan E J, Simoni A and Vitale B 1985 {\it Dynamical 
Systems. A Differential Geometric Approach to Symmetry and Reduction} 
(New York: Wiley)
\item\label{tre-grazie}
Choquet--Bruhat Y, De witt--Morette C and Dillard Bleick M 1977 {\it 
Analysis, Manifolds and Physics} (Amsterdam: North--Holland)
\item\label{logan}
Logan J D 1977 {\it Invariant Variational Principles} (New York: 
Academic Press)
\item\label{marek}
Demianski M, de Ritis R, Rubano C and Scudellaro P 1992 \pr {\bf D 46} 
1391 \\
Madsen M S, Mimoso P J, Butcher J A and Ellis G F R 1992 \pr {\bf D 46} 
1399
\item\label{infl}
Linde A D 1982 \pl {\bf 108B} 389 \\
Olive K A 1990 \prep {\bf 190} 307
\item\label{grad}
Gradshteyn I S and Ryzhik I M 1980 {\it Table of Integrals, Series,
and Products} (New York: Academic) 
\item\label{lucchin-mat}
Lucchin F and Matarrese S 1985 \pr {\bf D 32} 1316
\item\label{barrow}
Burd A B and Barrow J D 1988 \np {\bf B 308} 929
\item\label{moffat}
Moffat J W 1996 astro-ph 9603128 \\
Moffat J W 1996 astro-ph 9608202 
\end{enumerate}

\vfill

\end{document}